# Faster and Low Power Twin Precision Multiplier

V. Sreedeep, B. Ramkumar and Harish M Kittur

*Abstract-* In this work faster unsigned multiplication has been achieved by using a combination of High Performance Multiplication [HPM] column reduction technique and implementing a N-bit multiplier using 4 N/2-bit multipliers (recursive multiplication) and acceleration of the final addition using a hybrid adder. Low power has been achieved by using clock gating technique. Based on the proposed technique 16 and 32-bit multipliers are developed. The performance of the proposed multiplier is analyzed by evaluating the delay, area and power, with TCBNPHP 90 nm process technology on interconnect and layout using Cadence NC launch, RTL compiler and ENCOUNTER tools. The results show that the 32-bit proposed multiplier is as much as 22% faster, occupies only 3% more area and consumes 30% lesser power with respect to the recently reported twin precision multiplier.

## I. INTRODUCTION

In high performance digital systems such as microprocessors, FIR filters and digital signal processors etc., the multiplier is one of the key hardware blocks. So the design of multipliers stands challenging with advancement in technology. Many researchers have tried and are trying to design multipliers which offer either of the following- high speed, low power consumption, regularity of layout and hence less area or even combination of them, thereby making them suitable for various compact, low power and high speed VLSI implementations. However area and speed are two conflicting constraints. So, improving speed results in larger area and vice versa. Hence we try to find out the best trade off solution amongst them.

In recent trends the column compression multipliers are popular for faster computations due to their higher speeds [1-2]. The first column compression multiplier was introduced by Wallace in 1964 [3]. In 1965, Dadda altered the approach of Wallace by starting with the exact placement of the (3,2) counters and (2,2) counters in the maximum critical path delay of the multiplier [4]. In 2006, H. Eriksson along with his research team presented HPM reduction tree structure that has an ease of layout compared to Dadda's approach [5]. Compared to Dadda, HPM is slightly faster and consumes lesser power while area being the same. So we implemented the multiplier design using HPM.

The total delay of the multiplier can be split up into three parts: 1. The Partial Product Generation (PPG) 2. The Partial Product Summation Tree (PPST), and 3.The Final Adder [6]. Of these the dominant components of the multiplier delay are due to the PPST and the final adder. The relative delay due to the PPG is small. Therefore significant improvement in the speed of the multiplier can be achieved by reducing the delay in the PPST and the final adder stage of the multiplier. Here we are reducing the PPST delay using faster multiplication technique of performing the N-bit multiplication by '4' N/2-bit multiplications running in parallel and by the hybrid adder we are reducing the final adder delay.

The bit width of the multiplier is same as that of the bit width of the largest operand of the application that the processor executes. But most of the times the operands do not occupy the maximum width and utilizes the resources unnecessarily which results in power loss. In the year 2005 Magnus Sjalander explored on this idea to reduce this type of power consumption by using operand guarding technique and named it as Twin Precision Technique [6]. Now in this paper we are utilizing the same property for reducing the power and operator isolation is being performed using clock gating technique.

The remaining paper is organized as follows: Section II describes the design of the faster multiplier structure. Section III describes the design of hybrid adder and clock gating. Section IV is all about result analysis. Section V is the Conclusion. Section VI includes the bibliography.

*Assumtion-* **Multiplier bitwidth and multiplicand bitwidth are same.**

## II. DESIGN OF FASTER MULTIPLIER STRUCTURE

The first step of multiplication is PPG. PPG can be done by using an AND gate array or series of multiplexers. The next step is PPST. The PPG and PPST are shown in the following subsections

### A. Partial Product Generation [PPG]

Here we are considering N-bit multiplier, so let us assume

Multiplicand $Y = y_{n-1}\ y_{n-2}\ y_{n-3} \ldots \ldots \ldots y_3\ y_2\ y_1\ y_0$
Multiplier $\quad X = x_{n-1}\ x_{n-2}\ x_{n-3} \ldots \ldots \ldots x_3\ x_2\ x_1\ x_0$

So the partial products are $(y_j x_i) \in \{0,1\}$ where $i, j = 0,1,\ldots.n-1$. So for a N x N multiplication we are having a total of $N^2$ partial products as shown in the figure. 1(a). The value of $y_j x_i$ is '1' when both the operand bits are high and '0' when any one of the operand bit zero. Thus an AND gate can be used for the generation of partial products. For the convenience of representation of architecture we are considering N = 8.

Figure. 1(b) there are four different partial products arrays, of them the partial products that are marked in black are interdependent and cannot be used for parallel operation. But the partial products that are not in black can be operated in parallel as these are independent. This technique was used in [7] for twin precision multiplication.

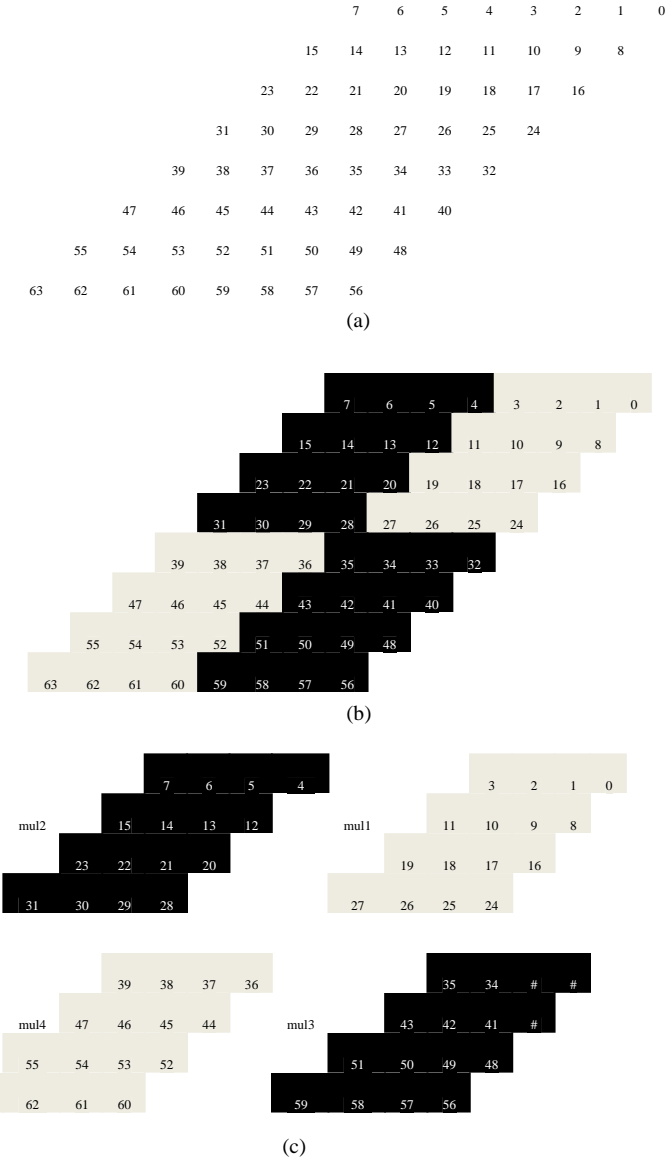

Figure. 1 Partitioning of partial products: (a) Partial product array for N = 8. (b) Partial Product array showing four partial product arrays of N = 4. (c) Rearranged partial products assigned to four different multipliers.

### B. The Partial Product Summation Tree [RPPST] for Recursive Multiplier

In our design we have segregated the partial products as shown in figure. 1(c) and each partial product array is given to a N/2-bit multiplier. Each N/2-bit multiplier uses HPM as column reduction technique [5] and uses ripple carry adder (RCA) as a final adder for computing the product. The four products thus obtained are used for the computation of final product. The proposed architecture with RCA as final adder and the flow of data is shown in the figure. 2. The architecture of each N/2- bit multiplier is shown in figure. 6 of [7].

Now as mentioned earlier the partial products that are dependent (marked in black in figure. 1(b) and 1(c)) are given to Multiplier M2 and M3 respectively. The products obtained from M2 and M3 are given to a N- bit RCA and the obtained result is given to N+1 bit RCA along with the MSB N/2 bits of product from M1 and the LSB N/2-bits of product from M4. The MSB N/2 bits of M4 product are given to N/2-1 bit RCA with '1' as carry input and calculating the result before the actual carry arrives. We have used a multiplexer for selecting the product based on the actual carry generated by N+1-bit RCA. This dependency and flow can be clearly observed in figure. 3.

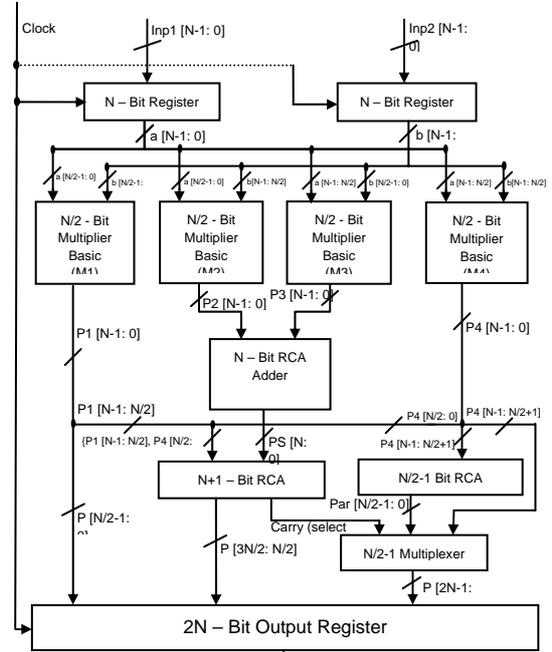

Figure. 2 Proposed architecture.

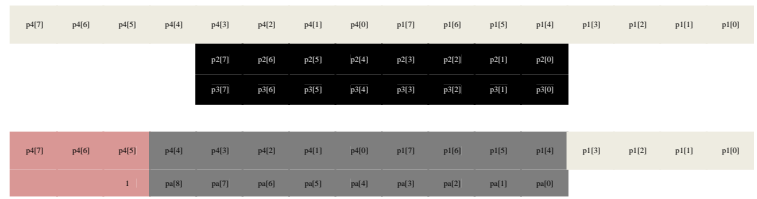

Figure. 3 Products of four multipliers [M1, M2, M3, M4]

### III. THE HYBRID FINAL ADDER DESIGN AND CLOCK GATING

#### A. MBEC ADDER DESIGN

In previous works the hybrid final adder designs used to achieve the faster performance in parallel multipliers were made up of Carry Look ahead Adder (CLA) and Carry Select Adder (CSLA) [8-10]. But CSLA occupies very large chip area than other adders (2x times compared to RCA). Here in this paper we are using MBEC (Multiplexers with Binary to Excess-1 Converters) to achieve the optimal performance. When compared to Carry Save Adder (CSA) and CLA adder MBEC is much faster and occupies lesser area and consumes less power compared to CSLA [11].

In the proposed architecture we have used N/2-1 bit RCA and multiplexer for adding the MSB n/2 bits of M4 product before the carry bit arrives by giving '1' as carry input as shown in figure.2 thus making the operation slower, occupying more area and consuming more power. Now the N/2-1 bit RCA is replaced with Binary to Excess-1 Converter (BEC). The logic diagram of a 5-bit BEC is shown in figure. 4. The BEC is used for further improving the speed.

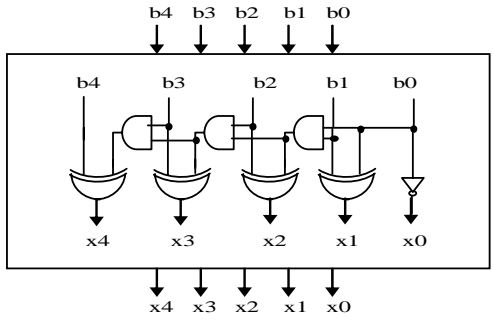

Figure. 4 5 Bit Binary to Excess-1 Converter

### B. Clock Gating

As mentioned earlier we are using operator isolation for the reduction of power using clock gating technique. Clock gating technique is nothing but to control the clock using one control signal. This can be performed using a simple AND gate. In our design previously we were using two N-bit registers for inputs are now replaced by 8 N/2-bit registers i.e., 2 for each multiplier which are driven by 3 different clocks generated by using the original clock and a control circuit as shown in figure. 5.

The circuit used here is a 2 to 3 decoder where our operation mode is input and we are generating 3 outputs that are in turn can be used for generating 3other clocks that control the flow of data in to the multipliers through registers. The decoder truth table is shown in Table I.

TABLE I
DECODER TRUTH TABLE

| Operation Mode | T[1] | T[2] | T[3] |
|---|---|---|---|
| 00 – Both M1 and M4 in operation for Twin Precision | 1 | 0 | 1 |
| 01 – Only M1 in operation | 1 | 0 | 0 |
| 10 – Only M4 in operation | 0 | 0 | 1 |
| 11 – Full Mode operation | 1 | 1 | 1 |

The advantage in this design compared to the regular twin precision multiplier in [7] is that we are isolating the operator instead of operand guarding. So in this design we can make use of one multiplier at a time for one N/2- bit multiplication but in regular twin precision we have to give all zeros for MSB N/2 bits of multiplier and multiplicand in order to operate the multiplier for same operation, so there is restriction in giving inputs which is not feasible always. But the control circuit here provides an advantage to overcome this.

The architecture shown in figure 5 not only increase speed but also provide the N/4 bit multiplication with less power consumption. This can be clearly observed from the result analysis.

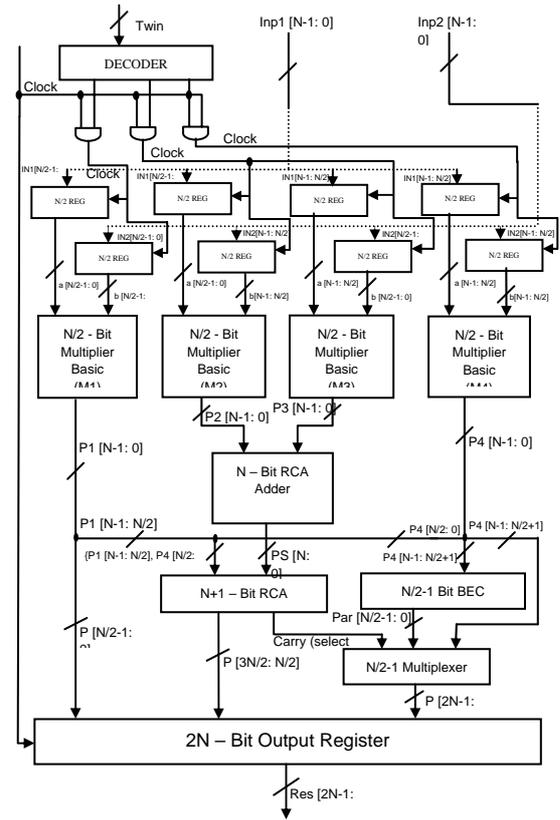

Figure. 5. Proposed Architecture with BEC Adder and Clock Gating

## IV. RESULT ANALYSIS

The comparison between the Table II (regular Twin precision multiplier in [7]) and Table III (proposed multiplier with BEC adder and clock gating) summarizes the enhanced performance of the proposed multiplier in terms of percentages which are listed in Table IV. The power results are calculated dynamically with 10000 inputs for 16 bit multiplier and with 15000 inputs for 32 bit multiplier. The summary of power comparisons in Table II and III for 16 and 32 bit are plotted respectively in figures 6 and 7. The area and timing comparison plots are shown in figures 8 and 9 respectively. The power delay products are shown in Table V.

TABLE II
REGULAR TWIN PRECISION MULTIPLIER (SJALANDER ET AL.)

| Multiplier Word Size | Type of Operation | Area (kµm2) | Time (ps) | Power (mW) |
|---|---|---|---|---|
| 16 x 16 | One 16 x 16 | 12.304 | 3 | 1.285 |
|  | Two 8 x 8 |  |  | 0.600 |
|  | One 8 x 8 |  |  | 0.325 |
| 32 x 32 | One 32 x 32 | 41.656 | 5.5 | 6.217 |
|  | Two 16 x 16 |  |  | 2.852 |
|  | One 16 x 16 |  |  | 1.507 |

TABLE III
PROPOSED MULTIPLIER

| Multiplier Word Size | Type of Operation | Area (kµm2) | Time (ps) | Power (mW) |
|---|---|---|---|---|
| 16 x 16 | One 16 x 16 | 12.471 | 2.6 | 1.331 |
|  | Two 8 x 8 |  |  | 0.568 |
|  | One 8 x 8 |  |  | 0.260 |
| 32 x 32 | One 32 x 32 | 42.985 | 4.25 | 4.362 |
|  | Two 16 x 16 |  |  | 1.846 |
|  | One 16 x 16 |  |  | 0.985 |

TABLE IV
PERFORMANCE OF THE PROPOSED MULTIPLIER WITH RESPECT TO REGULAR TWIN PRECISION MULTIPLIER

| Multiplier Word Size | Type of Operation | Area (%) | Time (%) | Power (%) |
|---|---|---|---|---|
| 16 x 16 | One 16 x 16 | +1.357 | -13.334 | 3.476 |
|  | Two 8 x 8 |  |  | -5.261 |
|  | One 8 x 8 |  |  | -19.927 |
| 32 x 32 | One 32 x 32 | +3.190 | -22.727 | -29.835 |
|  | Two 16 x 16 |  |  | -35.278 |
|  | One 16 x 16 |  |  | -34.618 |

TABLE V
POWER DELAY PRODUCT COMPARISON OF THE PROPOSED MULTIPLIER WITH RESPECT TO REGULAR TWIN PRECISION MULTIPLIER

| Multiplier Word Size | Multiplier | Energy (mJ) | Percentage (%) |
|---|---|---|---|
| 16 x 16 | Sjalander | 3.8484 |  |
|  | Proposed | 3.4593 | -10.1106 |
| 32 x 32 | Sjalander | 34.1971 |  |
|  | Proposed | 18.5397 | -45.7825 |

Figure. 10 represent all the percentage results shown in Table IV.

## V. CONCLUSION

We have successfully achieved a faster and low power multiplication by using a combination of High Performance Multiplication [HPM] column reduction technique and implementing a N-bit multiplier using 4 N/2-bit multipliers by rearranging partial products and acceleration of the final addition using a hybrid adder, low power has been achieved by using clock gating technique. The result analysis shows that area overheads are not significant when compared to the increase in speed and reduction in power consumption. The proposed multiplier design technique can be implemented with any type of parallel multipliers to achieve faster and low power performance. This work can be easily extended to signed multiplication.

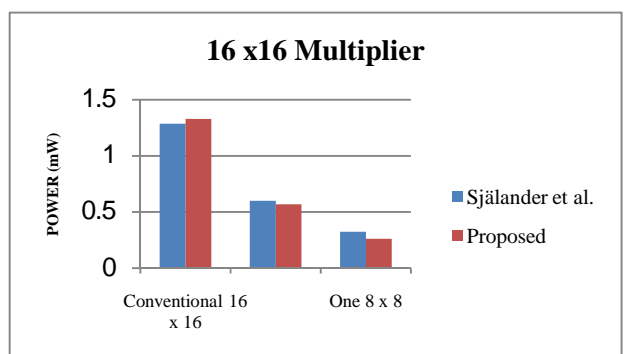

Figure. 6 Power Comparison Plot for 16 x 16 multiplier

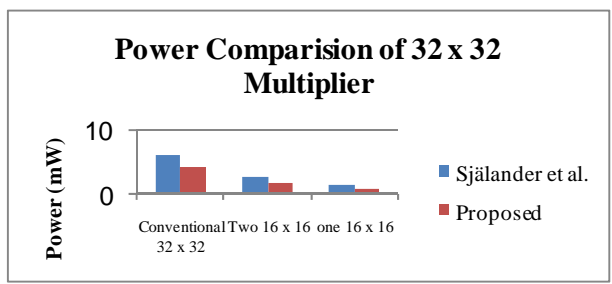

Figure. 7 Power Comparison Plot for 32 x 32 multiplier

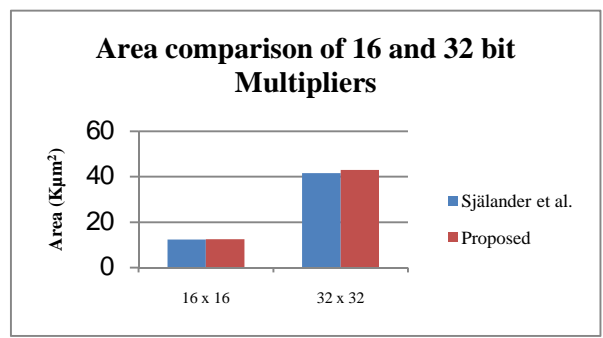

Figure. 8 Area Comparison Plot for both 16 x 16 and 32 x 32 multiplier

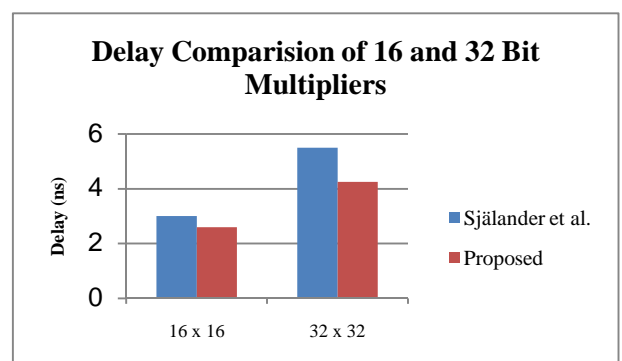

Figure. 9 Delay Comparison Plot for both 16 x 16 and 32 x 32 multiplier

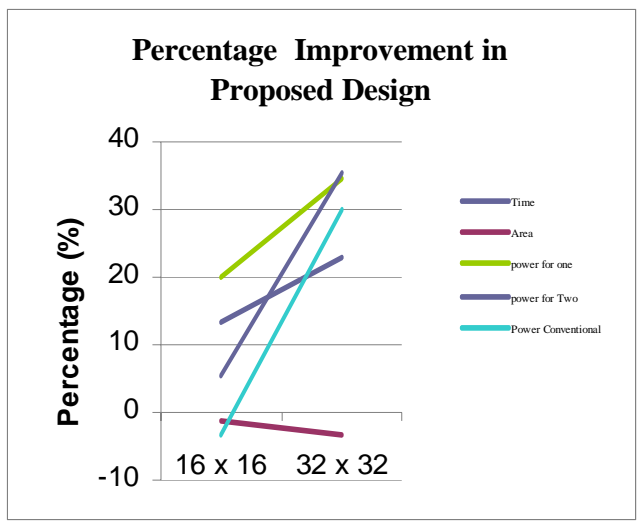

Figure. 10 Percentage Comparison Plot for Table I and Table II

## V. REFERENCES

[1] B.Parhami, "Computer Arithmetic", Oxford University Press, 2000.
[2] E. E. Swartzlander, Jr. and G. Goto, "Computer arithmetic," *The Computer Engineering Handbook*, V. G. Oklobdzija, ed., Boca Raton, FL: CRC Press, 2002.
[3] C. S. Wallace, "A Suggestion for a Fast Multiplier," *IEEE Transactions on Electronic Computers*, Vol. EC-13, pp. 14-17, 1964.
[4] Luigi Dadda, "Some Schemes for Parallel Multipliers," *Alta Frequenza*, Vol. 34, pp. 349-356, August 1965
[5] H. Eriksson, P. Larsson-Edefors, M. Sheeran, M. Själander, D. Johansson, and M. Schölin, "Multiplier reduction tree with logarithmic logic depth and regular connectivity," in *Proc. IEEE Int. Symp. Circuits Syst. (ISCAS), May 2006, pp. 4–8.*
[6] V. G. Oklobdzija and D.Villeger, "Improving Multiplier Design by Using Improved Column Compression Tree and Optimized Final Adder in CMOS Technology", *IEEE transactions on Very Large Scale Integration (VLSI) systems, Vol. 3, no. 2, June 1995.*
[7] Magnus Själander and Per Larsson-Edefors, " Multiplication Acceleration Through Twin Precision ", *IEEE Trans. O VLSI Systems vol. 17, no. 9, pp. 1233-1245 Sep 2009.*
[8] V. G. Oklobdzija and D.Villeger, "Improving Multiplier Design by Using Improved Column Compression Tree and Optimized Final Adder in CMOS Technology", *IEEE transactions on Very Large Scale Integration (VLSI) systems, Vol. 3, no. 2, June 1995.*
[9] Paul F.Stelling, "Design strategies for optimal hybrid final adders in parallel multiplier",Journal of VLSI signal processing, vol 14,pp,321-331,1996.
[10] Sabyasachi Das and Sunil P.Khatri,"Generation of the Optimal Bit-Width Topology of the Fast Hybrid Adder in a Parallel Multiplier", International Conference on Integrated Circuit Design and Technology (ICICDT) May, 2007.
[11] B.Ramkumar, Harish M Kittur and P.Mahesh Kannan, " ASIC Implementation of Modified Faster Carry Save Adder", *European Journal of Scientific Research, Vol. 42, Issue 1, 2010.*
[12] B.Ramkumar, Harish M Kittur, "Low Area, Low Power CSLA", IEEE Transactions on Very Large Scale Integration (VLSI) systems, accepted for publication DOI:10.1109/TVLSI.2010.2101621
[13] K.C. Bickerstaff, E.E. Swartzlander, M.J. Schulte, Analysis of column compression multipliers, Proceedings of 15th IEEE Symposium on Computer Arithmeitc,2001.
[14] W. J. Townsend, Earl E. Swartzlander and J.A. Abraham, "A comparison of Dadda and Wallace multiplier delays", Advanced Signal Processing Algorithms, Architectures and Implementations XIII. Proceedings of the SPIE, vol. 5205, 2003, pages 552-560.
[15] Danysh and Swamlander Jr., "A recursive fast multiplier", Asilomar Conf. **on** Signals, Systems & Computers, vol. *1,* pp. 197 -201, 1998.